\documentclass[a4wide,12pt]{article}
\usepackage{a4wide}
\usepackage[utf8]{inputenc}
\usepackage[english]{babel} 
\usepackage{url}

\usepackage{listings}
\usepackage{color}
\usepackage{graphicx}
\usepackage{fancyvrb}
\usepackage{amsmath}
\usepackage{wasysym}

\usepackage{tabularx,colortbl,booktabs}

\usepackage{pgf,tikz,pgfkeys,pgfplots}
\usetikzlibrary{shapes}
\usetikzlibrary{calc}
\usetikzlibrary{matrix}
\usetikzlibrary{positioning,shapes,shadows,arrows}

\date{\vfill Lausanne, June 30, 2014}

\title{\includegraphics[width=0.4\textwidth]{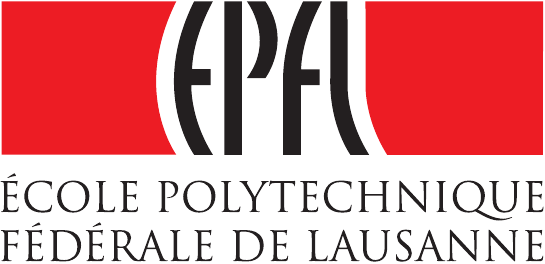} \\ \quad \\\quad \\\scshape Parallelisation of PyHEADTAIL \\ \quad \\ \large a Collective Beam Dynamics Code \\ for Particle Accelerator Physics}

\author{\quad \\ Adrian Oeftiger}

\definecolor{dkgreen}{rgb}{.13,.57,.33}
\definecolor{dkred}{rgb}{.57,.13,.33}

\begin{document}

\maketitle
\thispagestyle{empty}

\begin{abstract}
	The longitudinal tracking engine of the particle accelerator simulation application PyHEADTAIL shows a heavy potential for parallelisation. For basic beam circulation, the tracking functionality with the leap-frog algorithm is extracted and compared between a sequential C and a concurrent CUDA C API implementation for $10^6$ revolutions. Including the sequential data I/O in both versions, a pure speedup of up to $S=100$ is observed which is in the order of magnitude of what is expected from Amdahl's law. From $\mathcal{O}(100)$ macro-particles on the overhead of initialising the GPU CUDA device appears outweighed by the concurrent computations on the 448 available CUDA cores.
\end{abstract}

\pagebreak

\section{Introduction}

PyHEADTAIL is a program currently under development at CERN \cite{pyheadtailrep} designed for the simulation of the interaction of particle beams with electron clouds, the impact of impedance and space charge effects. It is used to study beam instabilities and emittance growth induced by the afore-mentioned collective effects. 

The code itself is written in python: various modules covering different physical aspects allow the user to choose a particular combination of what he wants to incorporate into his simulation script.

\begin{figure}[htp] \centering
	\includegraphics[width=0.5\linewidth]{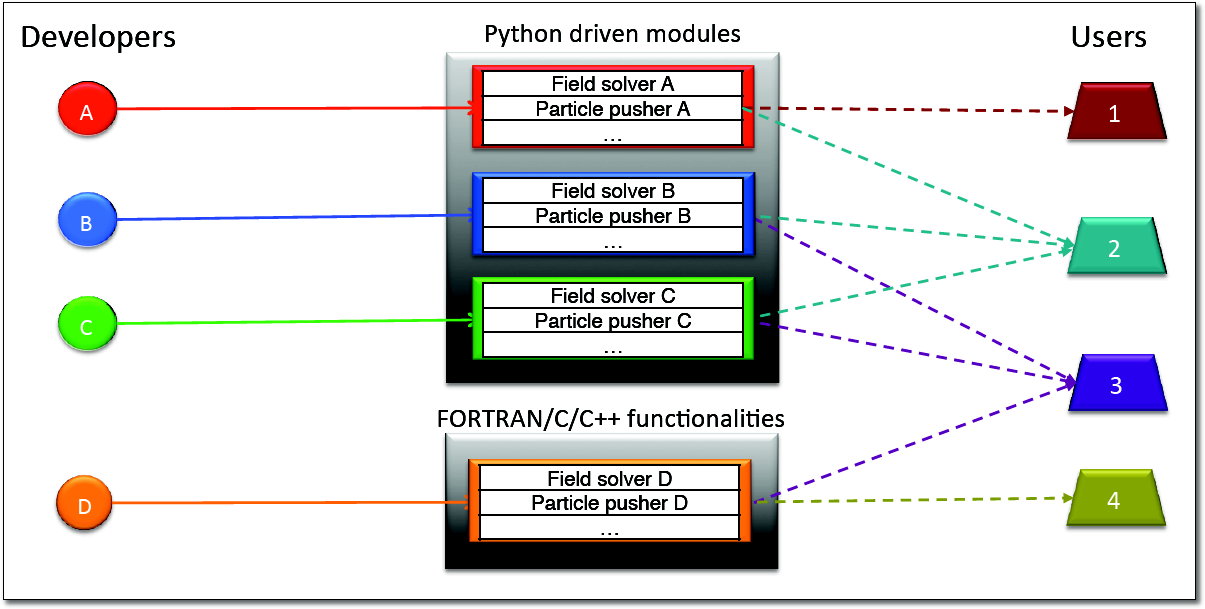}
	\caption{provided modules make it easier for users to set up their individual combination of covered physical aspects. \cite{vay}}
\end{figure}

PyHEADTAIL is a python port from the program HEADTAIL written in C and FORTRAN, which has been successfully employed in the world-wide beam dynamics community since at least 2002. Since PyHEADTAIL is still at an early stage of development, the current version does not exhibit any parallelised parts yet, it runs fully sequentially. This is where the actual project comes into play, it poses the starting point for future parallelisation efforts of PyHEADTAIL. 

\subsection{PyHEADTAIL's Tracking Engine}

\begin{figure}[h] \centering
	\includegraphics[width=0.4\linewidth]{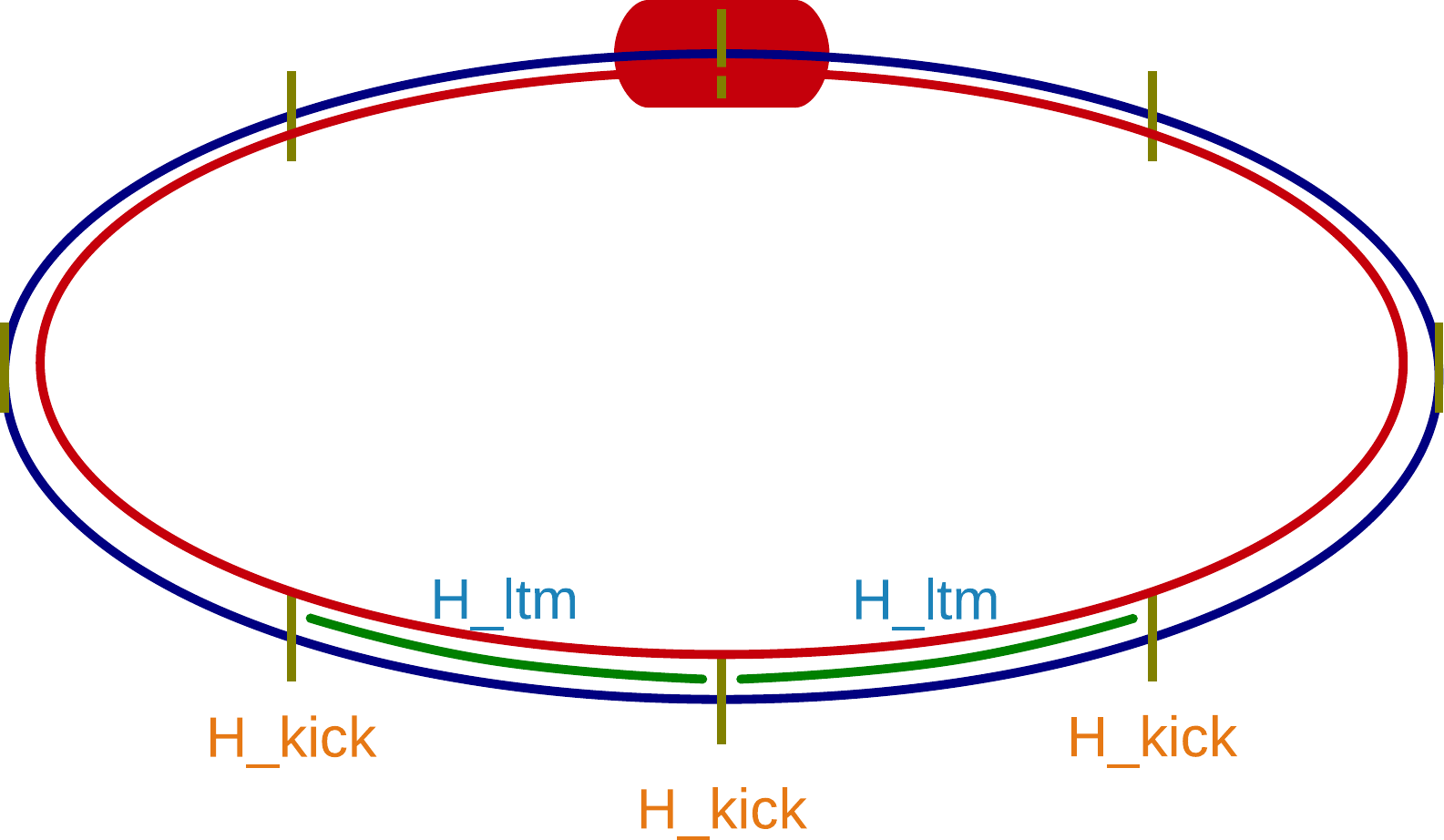}
	\caption{the particle distribution is pushed, resp.\ drifted (via the linear transfer matrix Hamiltonian) from one interaction point to another. \cite{klicobra}} \label{fig:pushing}
\end{figure}

PyHEADTAIL is a macro-particle simulation, i.e.\ the code pushes particles described by their 3D spatial positions and momenta from one interaction point to another, cf.\ figure \ref{fig:pushing}. At the interaction points -- depending on what type of interactions the user has included --, the fields of the particle distribution are constructed and act back on the particle distribution.

One essential part of PyHEADTAIL is the tracking engine which transports the macro-particles. One distinguishes between the transverse plane (perpendicular to the direction of motion within the accelerator ring) and the longitudinal direction (along the direction of motion).

\subsubsection{The Transverse Plane}

The macro-particle movement in the transverse plane is approximately described by linear transfer matrices, i.e.\ the same matrix is repeatedly acting on the typically $10^5$ to $10^7$ macro-particles' state vectors (to give an order of magnitude, this depends on the resolution that the physics requires to be noiselessly described and be properly reproduced). Currently this is done with sequential vectorised matrix multiplication on a single core. 

\subsubsection{The Longitudinal Plane}

The macro-particles' longitudinal motion is taken into account via various different integration algorithms, which allow for different levels of accuracy in exchange for computation time. To be more specific, the particles' motion in longitudinal phase space -- which comprises two coordinates, the spatial location $z$ and the corresponding (dimensionless) conjugate momentum $\delta$ -- is integrated turn by turn. Integration is performed via either the Euler-Cromer (first order), Leap-Frog (second order Verlet) or Runge-Kutta resp.\ Ruth (fourth order) numerical integration algorithms. These heavy calculations are independently done for each particle's entry in the $z$ and $\delta$ arrays, i.e. they are destined to be parallelised and outsourced to the GPU.

\section{Parallelisation of the Tracking Engine}

\subsection{Setup}
\subsubsection{Machine}

The machine on which the testing is conducted runs the linux distribution Ubuntu 12.04. Table \ref{tab:specs} summarises the most relevant machine parameters and specifications.

\begin{table}[htp] \centering \renewcommand{\arraystretch}{1.2}
\begin{tabular}{|r|l|}
	\hline \hline
	CPU & $2\times$ Intel Xeon E5-2630 \\ \hline
	CPU cores & $2\times 6$ \\ \hline
	RAM & 256 GB DDR3 \\ \hline
	CPU clock rate & 2.30 GHz \\ \hline
	CPU L3 cache & 15 MB \\ \hline
	instruction set & Intel AVX 64-bit \\ \hline\hline
	GPU & Intel Tesla C2075 \\ \hline
	GPU devices & 4 \\ \hline
	GPU DDR5 RAM & 5375 MB \\ \hline
	GPU clock rate & 1.15 GHz \\ \hline
	CUDA cores per device & $14\times 32 = 448$ \\ \hline
	max. no of threads per block & $1024\times 1024\times 64$ \\ \hline
	CUDA computing capability & 2.0 \\ 
	\hline \hline
\end{tabular}
\caption{Relevant Machine Specifications} \label{tab:specs}
\end{table}

The compilers used are \verb+gcc+ version 4.6.3 and \verb+nvcc+ 5.0 v0.2.1221. The used python version is 2.7.3.

\subsection{Approach}
In a typical longitudinal tracking setting as it occurs during my hollow bunch simulations \cite{hollow}, the tracking engine of PyHEADTAIL naturally plays the major role during the course of the simulation. A run with 100'000 macro-particles and 70'000 turns takes 1044.442 seconds on the above described setup. Neglecting the plotting part of the application, the python profiling tool \verb+cProfile+ reveals in figure \ref{fig:cprof} that the two core tracking functions that respectively update the $z$ and $\delta$ coordinate arrays of the particles consume 1039.034 seconds for the Leap-Frog algorithm, i.e. 99.5\% of the total time. The same run on my own desktop machine (exhibiting a CPU clock rate of 3.4 GHz) takes 707.164 seconds which perfectly scales with the clock rate ratio between the two machines.

\begin{figure}[htp]
\scriptsize
\begin{lstlisting}
Sun Jun 29 23:59:22 2014    hollow_bunches_stats.txt

         8023694 function calls (8019857 primitive calls) in 1044.442 seconds

   Ordered by: cumulative time
   List reduced from 1764 to 10 due to restriction <10>

   ncalls  tottime  percall  cumtime  percall filename:lineno(function)
        1    0.003    0.003 1044.444 1044.444 hollow_bunches.py:1(<module>)
        1    0.950    0.950 1044.209 1044.209 hollow_bunches.py:97(run)
    70000    0.698    0.000 1039.034    0.015 longitudinal_tracker.py:281(track)
   140000  619.140    0.004  629.595    0.004 longitudinal_tracker.py:126(track)
   140000   86.035    0.001  408.695    0.003 longitudinal_tracker.py:89(track)
   280001  222.015    0.001  225.573    0.001 beams.py:87(dp)
   419999   95.629    0.000   97.790    0.000 longitudinal_tracker.py:49(eta)
   980001    8.358    0.000    8.358    0.000 beams.py:66(beta)
   560002    2.693    0.000    7.973    0.000 beams.py:80(p0)
   140000    1.806    0.000    4.717    0.000 longitudinal_tracker.py:164(calc_phi_0)


\end{lstlisting}
\caption{cProfile output for \texttt{hollow\_bunches.py} .} \label{fig:cprof}
\end{figure}

In order to assess the speedup gain of a parallelisation of the longitudinal tracking engine quantitatively, its functionality is reproduced in an ANSI C code. The implemented physical situation corresponds to simple revolution tracking of the particles without complicated parameter changes of the accelerator setting (as in the above hollow bunch setting) -- the mechanism of which covers several hundred lines in the original python code but per se takes only a negligible part of the application running time.

A CUDA version of the sequential C code is written which performs the heavy computations in parallel for each particle on the GPU. 

Both the sequential C and the parallel CUDA version are compiled with aggressive optimisation flags \verb+-O2+. They both take as an input the same data file (generated by a python script) with two columns containing the coordinate values of $z$ and $\delta$ for various macro-particle numbers. This ensures that both perform the same calculations on exactly the same data. Reading in data files also resembles many applied simulation situations where real beam data are taken from measurements in the accelerator machines and then further processed in simulations.

Both the C and the CUDA version read in the file with the same sequential functions. Only the computational part is parallelised. The \verb+track+ function in the ANSI C code contains all of the tracking functionality. Some profiling reveals in figure \ref{fig:prof} that almost all the time is again spent on the tracking engine (and only negligibly on reading the data files).

\begin{figure}[htp]
\scriptsize
\begin{lstlisting}
$ ./long_tracking data10000.csv config2.cfg | head -8

INFO: file length evaluation time in ms:
  0.000000
INFO: file I/O evaluation time in ms:
 10.000000
INFO: track evaluation time in ms:
276190.031250
INFO: TOTAL evaluation time in ms:
276200.000000

\end{lstlisting}
\caption{Profiling output for \texttt{long\_tracking.c} given $10^6$ turns and $10^5$ macro-particles.} \label{fig:prof}
\end{figure}

So, finally we are all set to concentrate on the parallelisation results.

\subsection{CUDA: Parallelisation Design}

There are essentially two nested loops during the tracking: one going through all the turns and the other one going at each turn through all the particles in the two vectors \verb+z+ and \verb+dp+. Particles are independent of each other during the tracking itself, i.e. that loop can be parallelised. However, the turns necessarily need to be evaluated consecutively, i.e.\ they are sequential.

The instructions performed on each data entry of the two arrays are essentially the same each turn. Thus, the parallelisation design comprises the kernel knowing about the two arrays and performing the non-linear instructions concurrently on the data. Threads are spawned as many as there are particles resp.\ entries in the arrays.

Now there are two extremal choices -- put the loop over all turns into the kernel for the GPU threads or into the CPU code. The first one minimises memory access on the GPU (keyword: large latency threats!) and minimises the overhead created by spawning the threads. However, it turns out that for reasonably long total turn numbers (order of magnitude of 1 billion turns) the threads last too long, are considered to time out and are subsequently killed. (1 billion turns correspond to a 25h store of beam in the LHC, the current high energy collider at CERN.)

The opposite choice controls the loop over all turns from the CPU such that each GPU thread advances the corresponding particle by one turn. It is then important to \emph{synchronise} the spawning of threads as to yield to the strict time order of the integration. That results in very long idling times of single threads due to the comparatively large memory latencies on the GPU.

\begin{figure}[htp] \centering
	\includegraphics[width=0.8\linewidth]{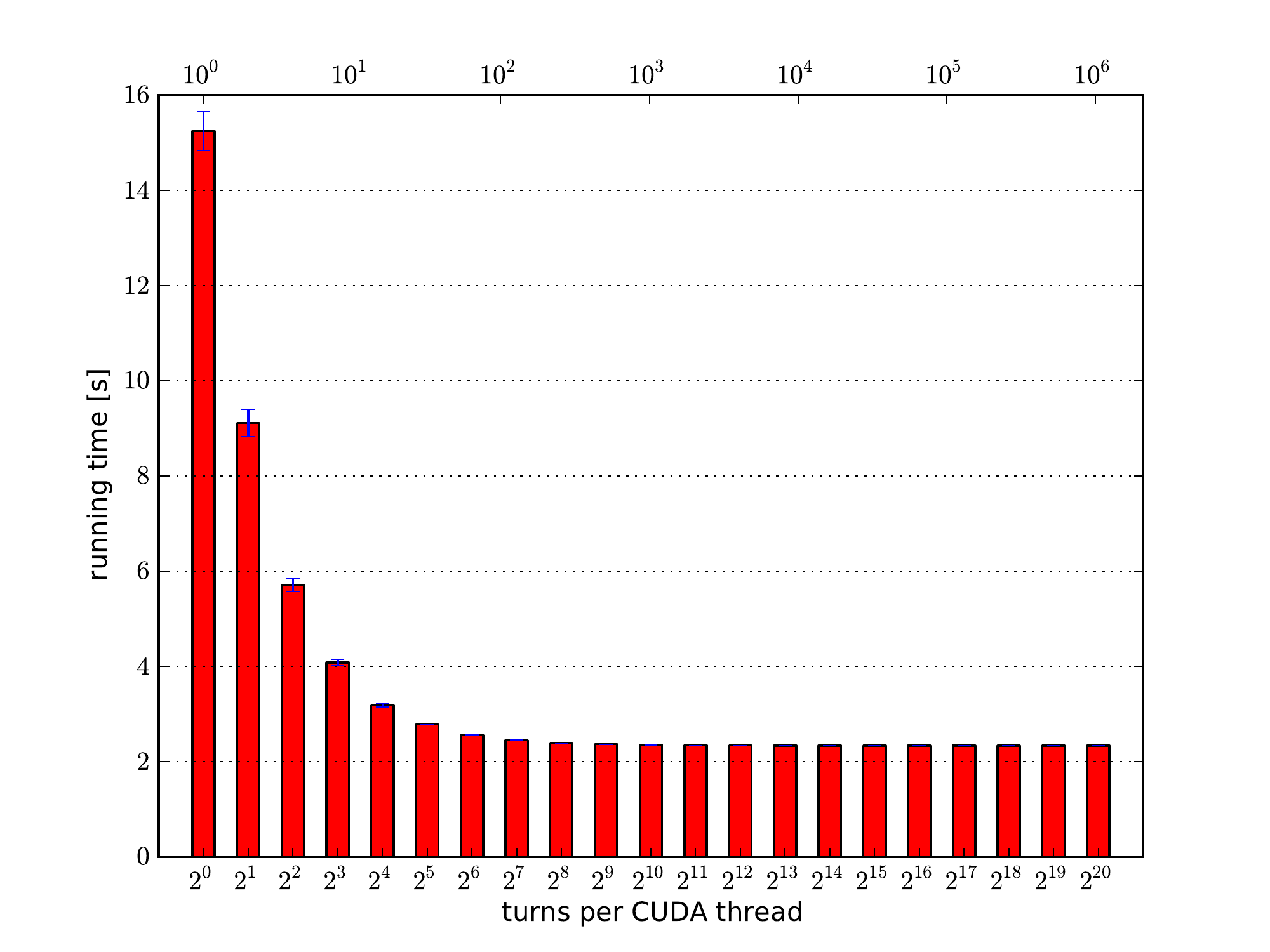}
	\caption{Tracking execution times for different number of turns per inner kernel loops, the overall number of turns is kept fixed at $10^6$ turns.}\label{fig:turnspercall}
\end{figure}

After these findings, a hybrid approach was chosen: a single thread contains a long enough loop to make use of the registers in order to save intermediate results at each turn and circumvent memory access, while its length is still short enough to prevent the thread to be killed. Another calling loop controlled from the CPU makes sure to cover the overall number of turns by enough threads steps.

Figure \ref{fig:turnspercall} shows the impact of the thread spawning overhead for low turns per thread, i.e.\ a high number of overall threads being spawned. A fixed number of $2^{20}\approx 10^6$ turns per inner kernel loop has then been established for the following evaluations.

\subsection{Speedup Evaluations}

\subsubsection{Theoretical Speedup}
Let us assess the speedup defined by
\begin{equation}
	S \doteq \frac{T_s}{T_p}
\end{equation}
(where $T_p$ is the parallelised code's execution time and $T_s$ the sequential code's execution time) that we expect to gain by parallelising the sequential ANSI C code.

Since in our case we have quite large data sets (the more macro-particles the longer the array and the better the resolution of a real bunch comprising $\approx 10^{12}$ particles) and we have quite some threads being spawned, we expect the real speedup to be oriented along the lines of Amdahl's law.

Amdahl's law states the strong scaling relation
\begin{equation}
	S_\text{Amdahl} = \frac{1}{(1-P)+\frac{P}{N}}
\end{equation}
for $S$ the ideal speedup, $P$ the fraction of the execution time of the parallelisable sequential code compared to the total execution time and $N$ the number of cores running the parallelised version, cf.\ \cite{nvidia}.
Thus, using one GPU device with 448 CUDA cores and taking $1-P=3.61\times 10^{-5}$ according to the profiling results of figure \ref{fig:prof} we obtain
\begin{equation}
	S_\text{Amdahl} = 441\pm 194 \quad ,
\end{equation}
which indicates the order of magnitude (since the time measurement uncertainty in figure \ref{fig:prof} -- we assume 1\permil\ uncertainty -- strongly affects the exact magnitude of this value).

NB: any overhead (such as copying data from CPU RAM to the GPU device RAM, different memory access times and especially initiating the cudaDevice etc.) is neglected in this calculation. Since the GPU runs at a different clock rate than the CPU, we have to take the ratio into account to get the real speedup:
\begin{equation}
	S = S_\text{Amdahl} \times \frac{1.15\text{ GHz}}{2.3\text{ GHz}} = 220\pm 97 \quad .
\end{equation}

\subsubsection{Measured speedup}
\begin{figure}[htp] \centering
	\includegraphics[width=0.7\textwidth]{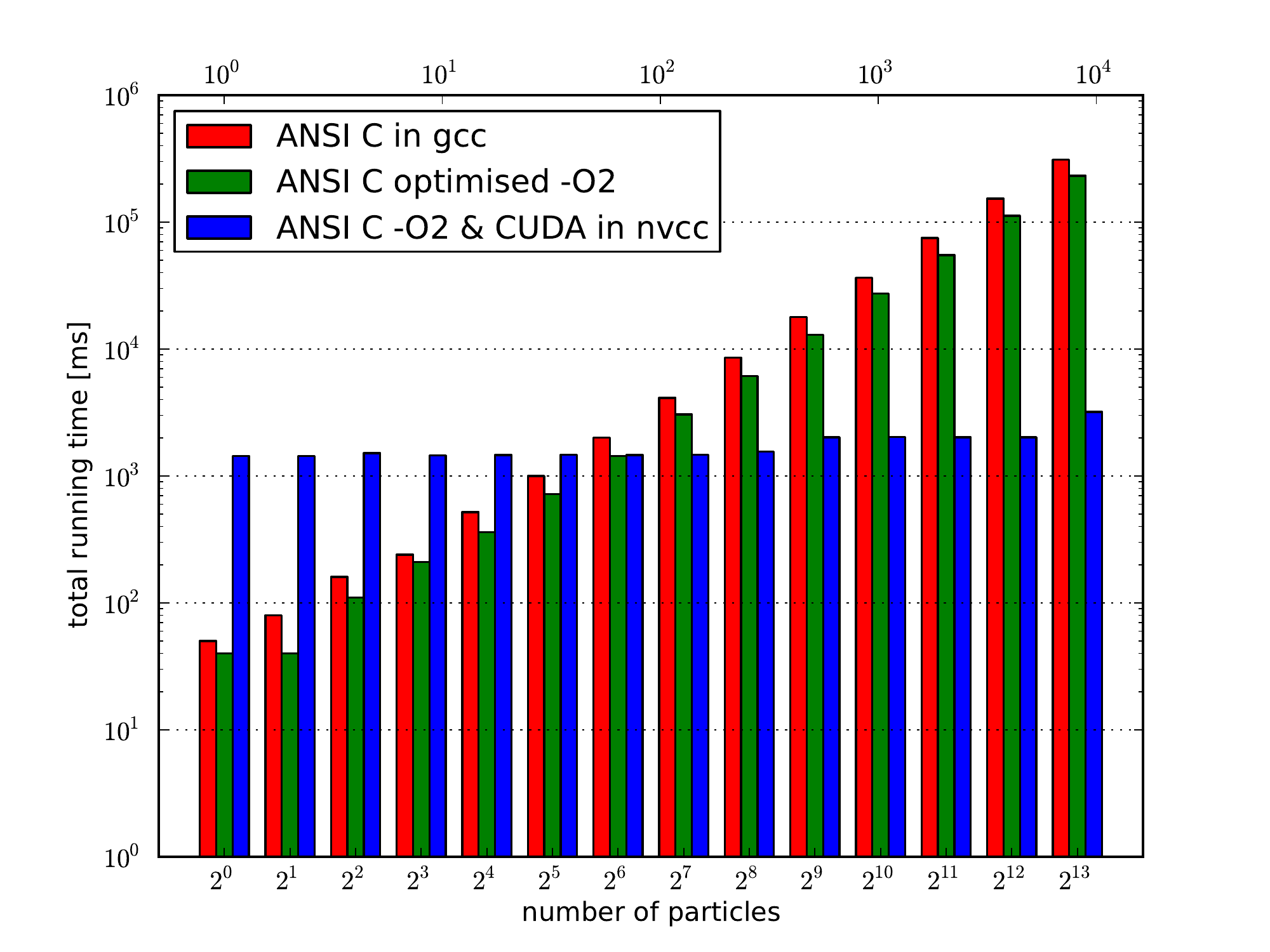}
	\caption{The total execution time including the file reading and the tracking for the sequential code compiled in debugging mode (red) and in aggressively optimised mode (green) as well as for the parallelised code compiled in aggressively optimised mode (blue).} \label{fig:eval}
\end{figure}

\begin{figure}[htp] \centering
	\includegraphics[width=0.7\textwidth]{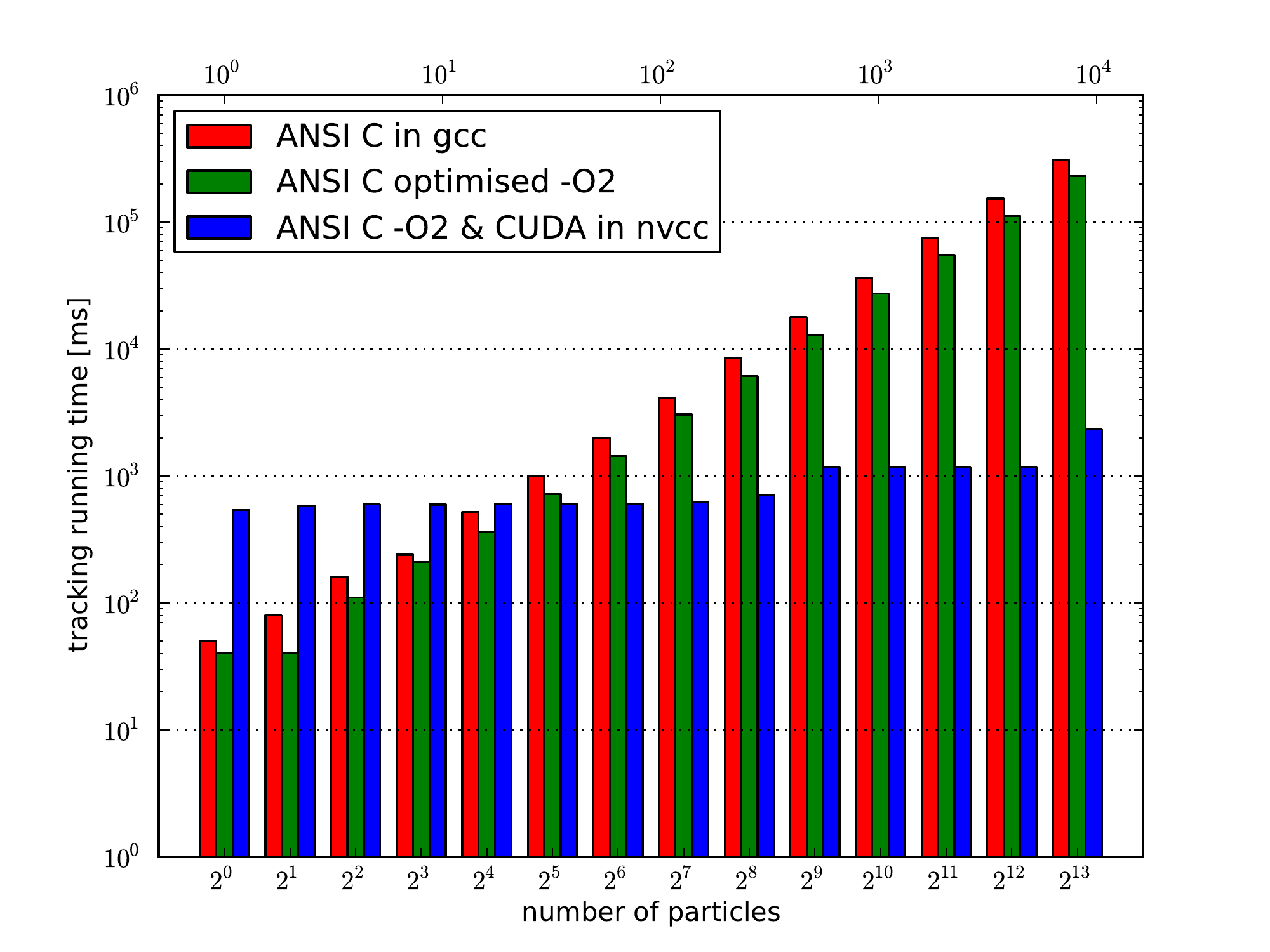}
	\caption{The kernel / tracking execution time for the sequential code compiled in debugging mode (red) and in aggressively optimised mode (green) as well as for the parallelised code compiled in aggressively optimised mode (blue).} \label{fig:eval2}
\end{figure}

We are using one-dimensional blocks holding 512 threads, i.e. we stay below the maximum of 1024 to have enough registers available for the kernel algorithm. To compare between a \verb+-O2+ flag optimised and non-optimised version of the sequential ANSI C code and the parallelised CUDA code, we are using data sets of 1 to 8192 macro-particles. Measurements of the execution times are shown in table \ref{tab:exec}. Figure \ref{fig:eval} compiles the total running times in a bar chart while the same situation is shown for the kernel / tracking running times only in figure \ref{fig:eval2}. All three versions essentially scale linearly with the number of turns given, as these are executed strictly sequentially. The CUDA GPU device initiation takes around $0.9$ seconds as seen in figure \ref{fig:cuprof} in the profiling of the parallelised CUDA C code (analogous to figure \ref{fig:prof}), so there is a considerable overhead. All CPU and overall times have been measured via the clock function of \texttt{time.h} which resembles the sole system CPU cycles spent on the program -- the final printing has been excluded from the measurements.

\begin{table}[htp] \centering \renewcommand{\arraystretch}{1.2}
\begin{tabular}{|c|c|c|c|c|c|c|c|c|c|}
	\hline\hline
	macro- & data file & \multicolumn{2}{c|}{ANSI C [ms]} & \multicolumn{2}{c|}{ANSI C \texttt{-O2} [ms]} & \multicolumn{2}{c|}{CUDA [ms]} & & \\ 
	particles & size & total & kernel & total & kernel & total & kernel & $S$ & $S_\text{pure}$ \\
	\hline\hline
1& 34 B& 50& 50& 40& 40& 1440& 541.3& 0.03& 0.07\\ \hline
2& 67 B& 80& 80& 40& 40& 1440& 581.6& 0.03& 0.07\\ \hline
4& 131 B& 160& 160& 110& 110& 1510& 598.3& 0.07& 0.18\\ \hline
8& 268 B& 240& 240& 210& 210& 1450& 597.1& 0.14& 0.35\\ \hline
16& 537 B& 520& 520& 360& 360& 1460& 603.3& 0.25& 0.6\\ \hline
32& 1.1 KB& 1000& 1000& 720& 720& 1470& 606.4& 0.49& 1.19\\ \hline
64& 2.1 KB& 2010& 2010& 1440& 1440& 1460& 606.3& 0.99& 2.38\\ \hline
128& 4.3 KB& 4130& 4130& 3050& 3050& 1470& 623.8& 2.07& 4.89\\ \hline
256& 8.4 KB& 8550& 8550& 6120& 6120& 1560& 707.3& 3.92& 8.65\\ \hline
512& 17 KB& 17890& 17890& 12920& 12920& 2020& 1167.0& 6.4& 11.07\\ \hline
1024& 34 KB& 36410& 36410& 27420& 27420& 2030& 1166.8& 13.51& 23.5\\ \hline
2048& 68 KB& 75000& 75000& 55090& 55090& 2020& 1166.5& 27.27& 47.23\\ \hline
4096& 134 KB& 153120& 153120& 112360& 112360& 2020& 1166.7& 55.62& 96.31\\ \hline
8192& 269 KB& 310300& 310290& 232320& 232310& 3200& 2332.3& 72.6& 99.6\\ 
	\hline\hline
\end{tabular}
\caption{Measured execution times for various code versions and number of macro-particles along with the final overall speedup $S$, the number of turns is kept fixed at $10^6$ turns.}\label{tab:exec}
\end{table}

\begin{figure}[htp]
\tiny
\begin{lstlisting}
$ ./long_tracking_cu data10000.csv config2.cfg | head -18

INFO: file length evaluation time in ms:
  0.000000
INFO: file I/O evaluation time in ms:
 10.000000
INFO: CUDA init evaluation time in ms:
890.000000
INFO: cudaMalloc evaluation time in ms:
  0.394176
INFO: cudaMemcpyHostToDevice evaluation time in ms:
  0.206752
INFO: Kernel evaluation time in ms:
2333.004150
INFO: cudaMemcpyDeviceToHost evaluation time in ms:
  0.191072
INFO: cudaFree evaluation time in ms:
  0.171168
INFO: TOTAL evaluation time in ms:
3280.000000

\end{lstlisting}
\caption{Profiling output for \texttt{long\_tracking.cu} given $10^6$ turns and $10^5$ macro-particles (analogous to figure \ref{fig:prof}).} \label{fig:cuprof}
\end{figure}

\newpage
\section{Conclusion}

Due to the overhead caused by the GPU CUDA device initiation, the use of the concurrent calculations on the GPU only pays off from a magnitude of $10^2$ to $10^3$ macro-particles on. As the usual setting in HEADTAIL / PyHEADTAIL is $10^5$ to $10^7$ macro-particles, a speedup of $S=\mathcal{O}(100)$ can be expected by using a parallelised version on the GPU. However, for more memory-bound applications such as calculating space charge effects between the particles or including impedance effects, the real speedup will be strongly affected by the large memory latency on the GPU. 

The next steps include porting the parallelised version into pyCUDA and integrating the parallelised tracking into PyHEADTAIL. For now, the strongly computation-bound tracking promises a large speedup, this will have to be analysed in greater detail for the more memory-bound physical effects that comprise the core part of (Py-)HEADTAIL and make it such a useful tool in the accelerator physics community.

\end{document}